# Do Automatic Comment Generation Techniques Fall Short? Exploring the Influence of Method Dependencies on Code Understanding

Md Mustakim Billah
mustakim.billah@usask.ca
University of Saskatchewan
Saskatoon, SK, Canada

Md Shamimur Rahman
mdr614@mail.usask.ca
University of Saskatchewan
Saskatoon, SK, Canada

Banani Roy
banani.roy@usask.ca
University of Saskatchewan
Saskatoon, SK, Canada

## Abstract

Method-level comments are critical for improving code comprehension and supporting software maintenance. With advancements in large language models (LLMs), automated comment generation has become a major research focus. However, existing approaches often overlook method dependencies, where one method relies on or calls others, affecting comment quality and code understandability. This study investigates the prevalence and impact of dependent methods in software projects and introduces a dependency-aware approach for method-level comment generation. Analyzing a dataset of 10 popular Java GitHub projects, we found that dependent methods account for 69.25% of all methods and exhibit higher engagement and change proneness compared to independent methods. Across 448K dependent and 199K independent methods, we observed that state-of-the-art fine-tuned models (e.g., CodeT5+, CodeBERT) struggle to generate comprehensive comments for dependent methods, a trend also reflected in LLM-based approaches like ASAP. To address this, we propose **HelpCOM**, a novel dependency-aware technique that incorporates helper method information to improve comment clarity, comprehensiveness, and relevance. Experiments show that HelpCOM outperforms baseline methods by 5.6% to 50.4% across syntactic (e.g., BLEU), semantic (e.g., SentenceBERT), and LLM-based evaluation metrics. A survey of 156 software practitioners further confirms that HelpCOM significantly improves the comprehensibility of code involving dependent methods, highlighting its potential to enhance documentation, maintainability, and developer productivity in large-scale systems.

## 1 Introduction

Program comprehension is a substantial part of software development, taking up to 58% of developers' time [84], with a code reading-to-writing ratio exceeding ten times to one [49]. Code comments are essential in this process, as studies have shown that commented code is significantly more understandable than code without comments [70, 80]. Despite their importance, developers often neglect to add or update comments [66], resulting in outdated [19, 78] or missing documentation [13, 64], which can hinder code comprehension, increase maintenance costs, and reduce overall software quality. Automatic code comment generation, also known as code summarization, helps bridge this gap by generating natural language descriptions of code, thereby enhancing comprehension and accelerating development and maintenance efforts [65].

Research on code comment generation has evolved significantly, moving from initial information retrieval techniques [15, 24, 59] to deep learning models [28, 32], fine-tuned pre-trained models [1, 16, 23, 76, 77], and, most recently, approaches using large language models (LLMs) [4, 20]. These techniques generate natural language summaries for code components, such as methods, to improve code comprehension. However, both fine-tuned and LLM-based techniques have two significant limitations. First, fine-tuning code-comment pairs alone often overlooks the contextual dependencies within the code, resulting in less relevant or generic comments. Second, LLM-based approaches heavily rely on finding structurally similar code examples within the training data. This can be challenging as each code method typically has unique functionality and structure, making such matches difficult to find and ultimately reducing the quality and relevance of the generated comments.

In Figure 1, we present a method written in Java, along with the original developer-written comment (i.e., ground truth) and comments generated by state-of-the-art (SOTA) baselines. While the ground truth provides detailed semantic information, accurately describing the method's behavior, viewing only the *close()* method in isolation makes it challenging to grasp these nuanced details. The baseline-generated comments, such as those by CodeT5+, CodeBERT, and ASAP, are overly generic, missing the depth needed to fully understand the method. In such cases, a developer would likely need to explore the implementation of the *awaitClose()* method to get the necessary context for understanding the actual functionality and intent behind *close()*. In this study, we classify the *close()* method as dependent because its functionality relies on another method, *awaitClose()*, which is considered independent (i.e., also termed as 'helper' methods) in this context. Inspired by similar examples presented in Figure 1, we hypothesize that generating accurate comments for dependent methods is challenging without access to the details of their associated helper methods.

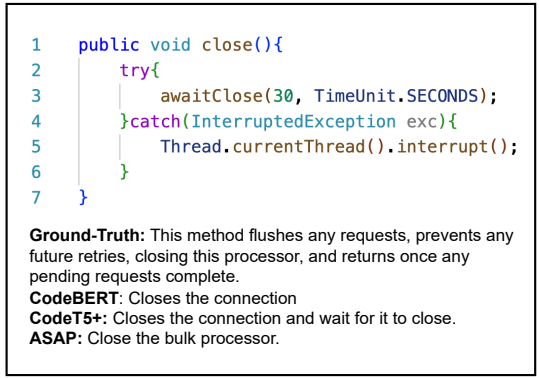

**Figure 1: Motivating example from a real-world project.**[1]

In this study, to test our hypothesis and address the research gaps, we propose **HelpCOM**, a novel comment generation technique designed to produce semantically rich comments for dependent methods while considering the details of associated helper methods. HelpCOM leverages GPT-4o[2], a SOTA large language model, to generate method-level comments by incorporating details from the helper methods linked to each dependent method. To evaluate HelpCOM's effectiveness, we created a dataset by extracting methods from 10 popular Java repositories on GitHub, linking dependent methods with their respective helper methods. We then compared the quality of comments generated by HelpCOM with those produced by baseline approaches. Our results indicate that HelpCOM outperforms all the baselines across standard evaluation metrics (BLEU [55], ROUGE-L [40], METEOR [5], SentenceBERT [57], Universal Sentence Encoder [10], CIDEr [72], SIDE [50]) for code summarization. Additionally, we validated the semantic quality of the generated comments through a survey of 156 experienced practitioners, including contributors to the GitHub repositories used in our study.

Our key contributions include the following:

- **Technique.** We introduce HelpCOM, a novel comment-generation technique that utilizes helper method information to generate highly comprehensive comments for dependent methods. To the best of our knowledge, this is the first approach of its kind.
- **Dataset.** We create a comprehensive dataset of dependent methods and their related helper methods, mined from popular GitHub repositories, which can be used in future studies on comment generation.
- **Comparative evaluation.** We compare different versions of HelpCOM with seven baseline comment generation techniques on our dataset, showing that HelpCOM significantly outperforms the baselines by 5.6% to 50.4% in the Overall Metric Score.
- **Developer's insight:** To further validate our findings and gather additional insights, we conducted a survey with 156 real-world software practitioners. The survey results demonstrate unanimous agreement among developers regarding the positive impact of our approach on code comprehensibility, particularly for dependent methods.
- **Data availability.** We provide a complete replication package[3] that includes the resources needed to reproduce our study, along with the results of our experiments.

## 2 Study Design

### 2.1 Goal

The goal of the study is described using the Goal-Question-Metric technique [9] as follows:

- Purpose: To enhance the comprehensibility
- Issue: Of generated comments for dependent methods
- Object: By utilizing helper methods
- Viewpoint: From the perspective of software practitioners

### 2.2 Research Questions

To achieve our study goal, we employ both qualitative and quantitative methods to explore the prevalence of dependent and independent methods. We also investigate the limitation of existing approaches for automatic comment generation in these scenarios, while proposing a novel solution to enhance comment quality and comprehensibility. Our analysis aims to address the following three Research Questions (RQs):

**RQ$_1$: To what extent does the evolution of dependent and independent methods vary in open-source projects?** Here, we investigate the relevance of dependent methods in software projects by comparing them to independent methods. Using metrics such as frequency, commit counts, and author involvement, we assess their engagement levels throughout project evolution. Higher engagement in either dependent or independent methods will highlight their importance as essential components of software projects, underscoring the need to prioritize them in method-level comment generation.

**RQ$_2$: How do existing approaches perform in generating comments for dependent methods compared to independent methods?** Recent works [4, 16, 20, 36, 76] have shown significant advancements in method-level comment generation. We utilize those techniques for both dependent and independent methods to assess whether these techniques face challenges in producing high-quality comments for dependent methods. Identifying any differences in comment quality will underscore the need for improved techniques tailored to dependent methods.

**RQ$_3$: To what extent does leveraging helper methods lead to the generation of higher-quality comments for dependent methods?** Prior studies [4, 20] have revealed that incorporating similar examples while generating comments for methods improves quality by providing additional context to LLMs. Building on this, we propose to investigate whether including information from helper methods can produce better quality comments for dependent methods compared to traditional approaches.

---
[1]https://bit.ly/4aMiddm
[2]https://openai.com/index/hello-gpt-4o/
[3]https://github.com/MustakimBillah/HelpCOM

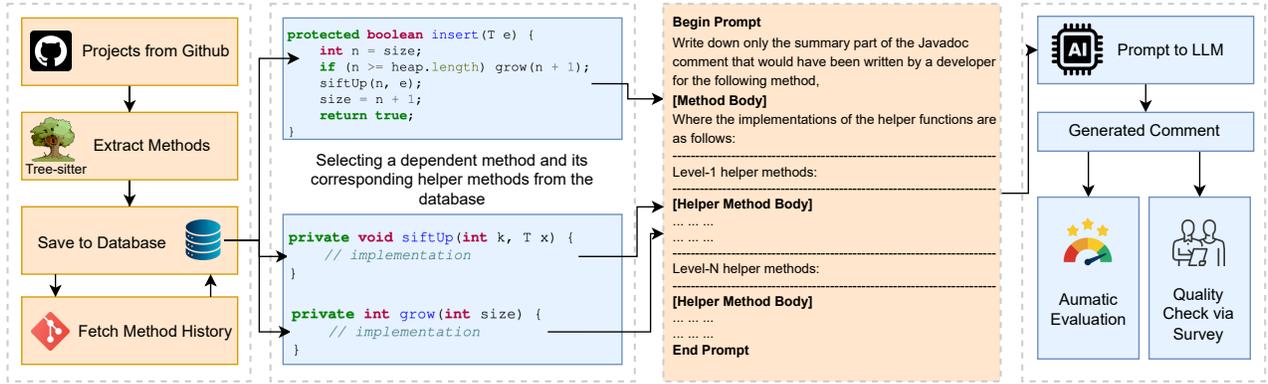

Figure 2: A step-by-step overview of our methodology.

### 2.3 Data Collection

Since the release of the CodeSearchNet dataset [31], it has been widely used in various code summarization studies [4, 20, 76]. This dataset contains two million pairs of code and corresponding comments from open-source libraries. However, CodeSearchNet, along with other datasets [30, 37], is unsuitable for our study because they lack information about helper methods called by dependent methods. For our research, we require a dataset that includes not only the code (method) and its corresponding comments but also details about the helper methods on which the code depends.

Preparing a new dataset tailored to our purpose posed a significant challenge. To address this, we mined the top 10 open-source Java projects on GitHub using SEART[4]. We applied strict inclusion criteria: i) a minimum of 10K stars, ii) at least 10K commits, iii) active pull requests (i.e., ensuring recent activity), and iv) at least 100 contributors. This filtering process enabled us to focus on projects with substantial activity and community engagement. An overview of the selected projects is provided in Table 1.

Table 1: Statistics about the mined Java repositories.

| Repository | Commits | Stars | Issues | PRs | Watchers | Contributors |
|---|---|---|---|---|---|---|
| elasticsearch | 76849 | 67687 | 34932 | 72998 | 2682 | 342 |
| netty | 11480 | 32937 | 6372 | 7522 | 1749 | 359 |
| logstash | 10745 | 14061 | 6698 | 9412 | 830 | 347 |
| spring-framework | 30253 | 55392 | 25328 | 4761 | 3346 | 368 |
| signal-android | 14670 | 25057 | 10715 | 2762 | 901 | 292 |
| presto | 22867 | 15667 | 6382 | 16414 | 861 | 328 |
| libgdx | 15364 | 22824 | 3722 | 3645 | 1160 | 332 |
| spring-boot | 49014 | 73268 | 34243 | 6257 | 3364 | 365 |
| druid | 14060 | 13237 | 4969 | 11537 | 589 | 358 |
| neo4j | 79919 | 12642 | 3630 | 9740 | 519 | 251 |

### 2.4 Dataset Preparation

*2.4.1 Method Extraction.* From each of the 10 projects, we aimed to extract all methods and their corresponding helper methods. We utilized Tree-sitter[5] to parse the projects, as it is a fast, robust, and dependency-free parser generator that efficiently builds and updates syntax trees for any programming language, even in the presence of syntax errors. First, we cloned the projects onto our local machine, selecting the 'main' or 'master' branch during cloning to ensure stable versions of the projects. We then parsed every '.java' file within each project. To extract methods from the parsed syntax tree, we used the *method_declaration* node type. Additionally, we extracted method-level comments using the *block_comment* node. By leveraging relevant node types, we also retrieved information such as the number of parameters and the start and end line numbers for each method. The extracted data was stored in MongoDB[6], a NoSQL[7] database, which stores data in JSON[8] format.

In total, we parsed 69,058 Java files and extracted 647,769 methods. We then focused on identifying helper methods, defined as methods explicitly implemented by developers (excluding built-in methods and self-calling recursive methods). A helper method is one that is called by another method, and the calling method is considered dependent on its helper methods. For example, if a method $M_1$ calls another method $M_2$ in its definition, then $M_2$ is the helper method of the dependent method $M_1$. Similarly, if $M_2$ calls another method $M_3$, then $M_3$ is the helper method of $M_2$. This chain of dependencies can extend further, forming a sequence $M_1 \rightarrow M_2 \rightarrow M_3 \rightarrow ....... \rightarrow M_n$. During extraction, we traced all such chains until no new helper methods were found.

To extract these method calls, we used the *method_invocation* node from the parse tree. After storing the method calls in our database, we matched these calls with the methods already stored in the database, using the method name and parameter count to retrieve the helper method bodies. Finally, we labeled methods without any associated helper methods as independent, while the rest were classified as dependent methods.

*2.4.2 History Fetching.* To analyze how dependent and independent methods evolve over time in software projects ($RQ_1$), we aimed to extract the change history of each method from GitHub. While there are several tools available for retrieving method histories from Git repositories, such as FinerGit [27] and CodeShovel [21], we opted to use Git[9] itself because of its accuracy and ease of use. For each method in our dataset, we used the *git log -L*

---

[4]https://seart-ghs.si.usi.ch/
[5]https://tree-sitter.github.io/tree-sitter/
[6]https://www.mongodb.com/
[7]https://cloud.google.com/discover/what-is-nosql
[8]https://www.json.org/json-en.html
[9]https://git-scm.com/

⟨start⟩,⟨end⟩:⟨file⟩[10] command. This command takes the start and end lines of a method along with the file path as parameters and returns the commit logs associated with that method. From the output, we extracted and recorded the total number of commits and authors associated with each method.

### 2.5 Baselines Selection

Fine-tuned pre-trained models have achieved SOTA performance in code comment generation, surpassing earlier techniques, as demonstrated by several studies [4, 16, 23, 76, 77]. In this work, we evaluate two SOTA models CodeT5+ [76] and CodeBERT [16] to assess differences in comment quality across various method types (RQ2). For CodeT5+, we used the 220M bimodal version from HuggingFace[11], fine-tuned on the CodeSearchNet dataset. This version was also employed by Wang et al. [76] for code summarization. For CodeBERT, we selected the polyglot version from Microsoft's GitHub repository[12], as Devanbu et al. [2] showed that multilingual fine-tuning enhances performance over monolingual models. We applied default parameters, adjusting source and target sequence lengths to 512 and 256, respectively, to align with the CodeT5+ configuration.

On the other hand, the emergence of LLMs enabled Ahmed et al. [4] to surpass pre-trained baselines with their LLM-based code comment generation technique, ASAP. ASAP employs *code-davinci-002*, a GPT-3 variant [8], as part of its methodology. However, since *code-davinci-002* was deprecated, we substituted it with OpenAI's recommended replacements: *gpt-3.5-turbo-instruct*[13]. We further experimented with ASAP using GPT-4o[2] to analyze its performance with a more capable model. We replicated ASAP by following their provided replication package[14] and included it as one of the baselines in our study. Additionally, we incorporated default GPT-4o, Llama-3.3[15], and CodeLlama [61] into our comparative analysis with HelpCOM.

### 2.6 Comment Generation through HelpCOM

The core concept behind HelpCOM is straightforward: leveraging helper methods to generate comments for a dependent method. We prompt LLMs to process this diverse information for their code understanding capabilities. For each dependent method in our dataset, we retrieve all of its associated helper methods and incorporate these into a custom-designed prompt. In designing this prompt, we followed OpenAI's guidelines[16] and took insights from related studies [4, 20, 52]. We then embed the dependent method and its helper methods into the prompt, as illustrated in Figure 2. We tried two variants of HelpCOM in our experiment, one of which only considers immediate helper methods of the parent method, and we term it as HelpCOM$_1$. The other one considers the total helper chain by taking into account every method. We term the latter one as HelpCOM$_N$.

We experimented with a number of LLMs to figure out the appropriate one for HelpCOM. For closed LLMs, we selected the GPT-4o for its GPT-4 Turbo-level performance in text generation, reasoning, and coding intelligence across standard benchmarks [11, 63]. GPT-4o also offers faster and more cost-effective processing. We set the model's temperature to 0.2 to reduce randomness and enhance the determinism of the generated comments, as per OpenAI API documentation[17]. Furthermore, we opted for codellama:7b-instruct and Llama3.3:70b as a representative of open LLMs, which we implemented following the Ollama[18] website. Finally, we pass the prompt to the model, retrieve the generated comment, and store it in our database.

### 2.7 Evaluation Metrics

To assess the effectiveness of existing techniques alongside our proposed method, we have used three types of evaluation metrics in this study.

*2.7.1 Syntactic Similarity Metrics.* These metrics calculate the textual similarity of the generated comments with the reference comments. BLEU [55], ROUGE-L [40], and METEOR [5] are among the most used syntactic similarity metrics in code summarization studies [6, 29, 32, 39, 52, 73, 75]. We use the smoothed BLEU-4 version [29] in our study. These metrics mostly calculate similarity by matching the n-gram overlapping of words. Each metric score ranges from 0 to 1, with higher values indicating stronger similarity between the generated and reference texts.

Though syntactic metrics are commonly used to evaluate generated comments, they have notable drawbacks. First, if the ground truth comments are of poor quality ( an issue documented in studies [18, 42, 78] ), better-generated comments may receive lower scores. Second, these metrics penalize summaries that are semantically equivalent but phrased differently, as the reference comment is just one possible way to comprehend the code. These limitations are well-recognized in software engineering research [26, 60].

*2.7.2 Semantic Similarity Metrics.* To overcome the shortcomings of the syntactic evaluation metrics, studies [10, 12, 46, 57, 72, 87] have proposed metrics that evaluate the quality of the generated comment against the reference comment by calculating the semantic similarity of the two. Haque et al. [26] reported in their study that cosine similarity of SentenceBert (SBERT) [57] and Universal Sentence Encoder (USEnc) [10] are closely comparable with the perception of human evaluators. The range of their score is between -1 to 1. We have also used CIDEr [72] and SIDE [50] metrics to report similarity scores in our study. Unlike other metrics, SIDE does not rely on reference comments; rather, it calculates the relevance of the generated summary by comparing it with the source code. The range of SIDE metric score is [-1, 1]. A higher score indicates closer similarity for these metrics.

*2.7.3 Evaluation by LLMs.* Recently, LLMs have adopted the role of the evaluator in several studies related to NLP [44, 74] and Software Research [3, 35, 67, 81]. LLMs can evaluate the quality of a generated comment in the absence of a reference summary and can lessen the dependency on human evaluation. Sun et al. [67] showed in their study that LLMs can evaluate the quality of the generated summaries of a method like human evaluators. They experimented with four LLMs and reported that GPT-4 has the closest correlation with human evaluation. In our study we also use GPT-4 and Llama-3.3. We followed exactly the same prompt used by Sun et al. [67] for the evaluation of comments with LLMs.

---

[10] https://git-scm.com/docs/git-log
[11] https://huggingface.co/Salesforce/codet5p-220m-bimodal
[12] https://github.com/microsoft/CodeXGLUE
[13] https://platform.openai.com/docs/deprecations#base-gpt-models
[14] https://zenodo.org/records/10494170
[15] https://ollama.com/library/llama3.3
[16] https://platform.openai.com/docs/guides/prompt-engineering
[17] https://bit.ly/42KZGfr
[18] https://ollama.com/

*2.7.4 Overall Metric Score.* To report the findings of our study, we defined an overall metric score (OMS) by combining all the metrics. We calculate the average scores of Syntactic ($Syn_{avg}$), Semantic ($Sem_{avg}$), and LLM-based ($LLM_{avg}$) evaluation metrics and formulate the following two equations: OMS excluding LLM-based metrics ($OMS_{ss}$), and OMS including LLM-based metrics ($OMS_{ssl}$).

$$OMS_{ss} = (0.46 \times Syn_{avg}) + (0.54 \times Sem_{avg}) \quad (1)$$

$$OMS_{ssl} = (0.30 \times Syn_{avg}) + (0.35 \times Sem_{avg}) + (0.35 \times LLM_{avg}) \quad (2)$$

We maintained the weight ratio of $Syn_{avg}$ to $Sem_{avg}$ as 30:35 in both equations (1) and (2). We assigned slightly lower weights to syntactic metrics compared to the other two metrics due to their limitations discussed in section 2.7.1. In this study, we report the scores for all evaluation metrics on a scale of 100.

## 2.8 Survey Overview

This survey aims to gather insights from software practitioners on how method-level comments and helper method information affect source code comprehensibility. The survey and questionnaires were approved by the Ethics Behavioral Board of the University of *Anonymous*, and all materials, including the complete survey, are available in the replication package. The survey is divided into several sections:

*Demographics.* The survey guarantees complete anonymity, and no personally identifiable information is requested except for the participant's role, programming experience (in years), and current job location (country). Participants are also asked for their consent to participate in the survey.

*Survey Tasks.* We designed three tasks categorized by difficulty level: easy, medium, and complex. Each task is divided into three interconnected subtasks: i) **Initial evaluation:** participants were provided with a Java method that calls one or more helper methods. However, the definitions of these helper methods were not included. Participants were asked to analyze the method and rate its difficulty of understanding on a scale of 1 to 10, where a lower score indicates easier comprehension and a higher score indicates greater difficulty; ii) **Re-evaluation with helper methods:** here, participants were provided with the definitions of all helper methods called by the original method. They were then asked to re-evaluate and rate the method's difficulty of understanding using the same scale; iii) **Comment selection:** finally, participants were presented with three method-level comments generated by selected baselines (i.e., CodeT5+, and ASAP) and the proposed HelpCOM. Without knowing which comment originates from which model (to avoid bias), participants must select the comment they find most comprehensible for the given method, without access to the helper methods' definitions. After selecting the comment, participants indicated whether the chosen comment helped them understand the method more comprehensively than the other given comments.

*Gather Feedback.* At the end, we gathered participant feedback on the usefulness of well-written code summaries and their potential impact on code comprehension. Participants were asked whether they agreed that code summaries improve understanding and whether they would be interested in a future tool that considers method dependencies in summary generation. Additionally, we invited them to share any concerns or suggestions for improving the approach.

*Participants Recruitment.* Our survey aims to engage practitioners with expertise in programming and software development knowledge. To recruit participants, we contacted via LinkedIn and used 6,000 available contributor email addresses obtained from the 10 subject systems selected in this study. Notably, we refrained from filtering email addresses to minimize sampling bias and sent invitations to all publicly available email addresses. Participation is voluntary, allowing individuals to decide whether to participate in the study. We also encouraged participants to forward the invitation to those who meet the selection criteria, ensuring broader coverage. Strict confidentiality measures were implemented to ensure participant privacy and data protection.

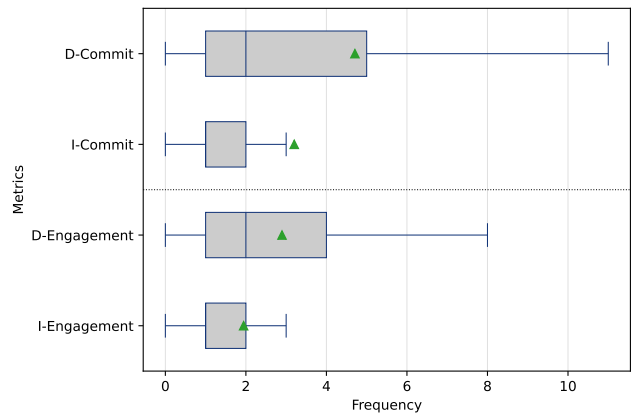

**Figure 3: Frequency of commit and engagement by method types. (▲) indicates the mean value of each distribution.**

## 3 Results

### 3.1 RQ$_1$: Evolution of Methods

Among the 647,769 methods in our dataset, 448,572 (69.25%) are dependent, and 199,197 (30.75%) are independent, indicating that dependent methods are predominant in software projects. Analyzing the method histories, we observe that 77.37% of the 2,699,964 commits were related to dependent methods, while 22.63% were for independent methods. To further understand engagement with these methods, we analyzed the number of authors associated with each method, as a higher number of authors per method suggests greater engagement with the codebase.

Figure 3 presents box plots comparing the frequency of commits and engagement (authors) for dependent methods (D-Commit, D-Engagement) versus independent methods (I-Commit, I-Engagement) across the 10 projects in our dataset. Dependent methods show higher engagement and change frequency, as reflected by their elevated Q3 (third quartile) and maximum values. Additionally, dependent methods have higher mean values for commits and engagement, at 4.7 and 2.6, respectively, compared to 3.1 and 1.9 for independent methods. These findings suggest that dependent methods are more actively modified and require greater comprehensibility, as changes to these methods can impact multiple parts of the codebase. Since dependent methods are more prone to change, they demand clear and informative comments to assist developers in understanding their evolving functionality. Therefore, comment-generation techniques should prioritize dependent methods to improve code comprehensibility and maintainability.

> **Summary RQ$_1$:** Dependent methods dominate in open-source projects, making up 69.25% of methods and 77.37% of commits. They experience higher engagement and more frequent changes, with greater mean values for both commits and engagement than independent methods, suggesting they attract more developer attention and should be prioritized in comment generation.

## 3.2 RQ$_2$: Performance Analysis of the Baselines

Among the methods in our dataset, 75,876 have method-level comments written by developers, with 41,583 categorized as dependent and 34,293 as independent. We used baselines CodeT5+ and CodeBERT to generate comments for these methods, and Table 2 presents their performance across the evaluation metrics.

**Table 2: Performance of baselines on the entire dataset.**

| Evaluation Metrics | | CodeT5+ | CodeBert |
|---|---|---|---|
| Syntactic Similarity | BLEU | 2.3 | 2.8 |
| | METEOR | 12.8 | 20.4 |
| | ROUGE-L | 13.4 | 17.6 |
| Semantic Similarity | CIDEr | 6.7 | 8.0 |
| | SBERT | 50.9 | 49.8 |
| | USEnc | 36.7 | 39.6 |
| | SIDE | 86.0 | 85.0 |
| OMS$_{ss}$ | | 28.7 | 30.9 |

The performance of these baselines was not satisfactory both on syntactic and semantic metrics. Ahmed et al. [4] reported BLEU scores of 19.83 and 20.22 for CodeT5+ and CodeBERT on a test sample of 1,000 Java methods. Sun et al. [67] also reported higher values on semantic metrics for Java methods for the lowest-performing model (for example, 59.47 in SentenceBERT) compared to the values of Table 2.

This discrepancy could be due to various factors; for example, some ground-truth comments in our dataset might lack clarity or consistency, introducing noise and impacting metric results, or the baselines might not generalize well to diverse datasets. To address this, we applied the SIDE metric proposed by Mastropaolo et al. [50], which scores the alignment between generated summaries and code snippets on a scale from 0 to 1. Using a SIDE threshold of >= 0.8, we refined our dataset to 48,260 methods (28,240 dependent and 20,020 independent).

As baseline ASAP [4] and variant of HelpCOM use the paid GPT API in their methodologies, we further sampled our dataset at a 95% confidence level with a 5% margin of error, resulting in 380 methods each for dependent and independent categories, totaling 760 methods. We then ran our three baselines on these sampled methods and presented the results in Table 3. We can observe that the OMS$_{ss}$ scores for CodeT5+ and CodeBert have been improved than the OMS$_{ss}$ scores of Table 2, indicating the quality of our sample dataset.

**Table 3: Performance comparison of baselines over Independent (I) and Dependent (D) methods.**

| Metrics | | CodeT5+ | | CodeBERT | | ASAP | |
|---|---|---|---|---|---|---|---|
| | | I | D | I | D | I | D |
| Syntactic Similarity | BLEU | 7.4 | 2.6 | 5.4 | 2.8 | 12.7 | 6.4 |
| | METEOR | 26.7 | 16.6 | 28.9 | 18.9 | 36.3 | 24.6 |
| | ROUGE-L | 29.5 | 19.4 | 24.9 | 18.7 | 35.6 | 26.0 |
| Semantic Similarity | CIDEr | 13.3 | 9.4 | 12.9 | 9.2 | 18.6 | 13.5 |
| | SBERT | 61.4 | 54.8 | 59.8 | 54.9 | 65.8 | 61.4 |
| | USEnc | 52.7 | 46.0 | 48.9 | 42.9 | 58.3 | 54.9 |
| | SIDE | 91.2 | 90.1 | 89.8 | 88.9 | 91.5 | 91.9 |
| LLM Based | GPT-4 | 64.0 | 67.2 | 56.4 | 54.2 | 90.6 | 92.8 |
| | Llama-3.3 | 67.2 | 68.6 | 64.2 | 53.8 | 89.4 | 91.2 |
| OMS$_{ss}$ | | 39.3 | 32.9 | 37.6 | 32.6 | 44.6 | 38.7 |
| OMS$_{ssl}$ | | 48.5 | 45.2 | 45.5 | 40.1 | 60.5 | 57.3 |

The results clearly show that all baseline methods achieve higher scores on independent methods than on dependent ones across the majority of the metrics (including OMS$_{ss}$ and OMS$_{ss}$). More specifically, overall metric scores OMS$_{ssl}$ were 7.3% to 13.5% higher for Independent methods than dependent ones across the baselines. To confirm if this difference is statistically significant, we conducted a two-sided Mann-Whitney U Test [47] over the Syntactic and Semantic similarity metrics, as the samples of independent and dependent methods are not identical. Except for values of SIDE on CodeBert and ASAP, we found that in each case, the difference is significant (p-value < 0.05), indicating that baseline methods face greater challenges in generating quality comments for dependent methods than for independent ones.

> **Summary RQ$_2$:** Our analysis shows that baseline models (CodeT5+, CodeBERT, and ASAP) perform significantly better (ranging from 7.3% to 13.5%) on independent methods than dependent ones, with most of the metrics favoring independent methods. Mann-Whitney U test confirmed this difference as statistically significant, indicating challenges for these models in generating high-quality comments for dependent methods.

## 3.3 RQ$_3$: Comparing HelpCOM with Baselines

Given the relatively low metric scores for dependent methods in our sample compared to independent methods (TABLE 3), we aimed to reduce noise to avoid under-reporting baseline performance. Developers of the selected open-source projects are familiar with their respective codebases. As a result, their method-level comments may be less comprehensive for general practitioners, such as novice programmers or new contributors, who lack prior understanding. To assess the general comprehensibility of ground-truth

Table 4: Performance comparison of HelpCOM with baselines.

| Approaches | Syntactic Metrics | | | | Semantic Metrics | | | LLM Evaluation | | Overall Score | |
|---|---|---|---|---|---|---|---|---|---|---|---|
| | BLEU | METEOR | ROUGE-L | CIDEr | SBERT | USEnc | SIDE | Llama-3.3 | GPT-4 | $OMS_{ss}$ | $OMS_{ssl}$ |
| CodeT5+ | *4.83 | *24.03 | *22.39 | *10.01 | *56.71 | *49.27 | *90.14 | *54.60 | *58.20 | 35.69 | 42.90 |
| CodeBERT | *3.59 | *20.88 | *20.55 | *8.28 | *54.30 | *43.81 | *90.34 | *46.0 | *45.80 | 33.46 | 37.78 |
| ASAP | 7.07 | 29.81 | **28.11** | 11.42 | 61.40 | **55.81** | 92.29 | *80.20 | *79.80 | 39.79 | 53.83 |
| ASAP (GPT-4o) | *3.10 | *15.19 | *13.45 | *6.51 | *36.29 | *16.20 | *68.88 | *31.60 | *31.20 | 22.13 | 25.35 |
| GPT-4o | 7.15 | *28.99 | 26.86 | *10.80 | 61.54 | 54.52 | 92.95 | 85.60 | 86.80 | 39.33 | 55.70 |
| CodeLlama | *3.08 | *25.88 | *16.59 | *7.69 | *55.88 | *42.48 | *91.08 | *77.80 | *83.20 | 33.60 | 49.98 |
| Llama-3.3 | *6.57 | 29.89 | 26.05 | *10.25 | 61.97 | 54.57 | 92.33 | 86.40 | 86.60 | 39.17 | 55.70 |
| $HelpCOM_1$ (GPT-4o) | 7.47 | 30.96 | 26.83 | 11.42 | 62.36 | 55.20 | 92.75 | 86.00 | 86.40 | 39.94 | 56.10 |
| $HelpCOM_1$ (Llama-3.3) | 6.92 | 30.57 | 26.60 | 10.87 | 62.41 | 55.56 | 92.18 | *82.20 | *85.40 | 39.66 | 55.08 |
| $HelpCOM_1$ (codeLlama) | *4.98 | *26.99 | *20.24 | *9.86 | *57.69 | *47.06 | *89.64 | *69.80 | *74.40 | 35.58 | 48.33 |
| $HelpCOM_N$ (GPT-4o) | **8.30** | **31.39** | 27.31 | **12.03** | **62.81** | 54.94 | **92.98** | **87.20** | **87.80** | **40.35** | **56.82** |
| $HelpCOM_N$ (Llama-3.3) | 7.26 | 30.80 | 26.44 | 11.41 | 62.28 | 55.54 | 92.22 | *80.40 | *82.60 | 39.79 | 54.35 |
| $HelpCOM_N$ (codeLlama) | *4.19 | *26.07 | *19.78 | *9.24 | *56.89 | *46.10 | *88.06 | *66.60 | *74.40 | 34.71 | 47.20 |

Note: * marked cells indicate significant differences ($p < 0.05$) in performance between $HelpCOM_N$ (GPT-4o) and the corresponding approach in the same row, based on the metric in the respective column, determined by the Wilcoxon signed-rank test.

comments written by developers, three professional Java developers, one with over eight years of experience and two with over three years, voluntarily participated, independently rating 380 ground-truth comments from the dependent sample as either 'acceptable' or 'unacceptable' based on semantic accuracy. After reaching an agreement with a Fleiss' Kappa [17] score of 83.6% (above the 70% threshold), the most senior developer revised the unacceptable comments to improve clarity and accuracy. To eliminate bias, the study goals were kept hidden from the raters.

We then ran $HelpCOM_1$ and $HelpCOM_N$ alongside the baseline models on the refined dependent sample set. We experimented with three LLMs, GPT-4o, CodeLlama, and Llama-3.3, for each variant of HelpCOM. Additionally, we experimented with all the LLMs used for HelpCOM separately to analyze their performance with and without HelpCOM. TABLE 4 shows the results of all the comment generation approaches over the evaluation metrics used in our study. The highest values for each metric are highlighted in bold.

Among the 13 approaches, $HelpCOM_N$ (GPT-4o) achieved the highest overall metric scores, with an $OMS_{ss}$ score of 40.35 and an $OMS_{ssl}$ score of 56.82. Specifically, $HelpCOM_N$ (GPT-4o) showed an overall improvement in the $OMS_{ssl}$ score ranging from 5.6% to 50.4% compared to the baselines (CodeT5+, CodeBERT, ASAP). Additionally, $HelpCOM_N$ (Llama-3.3) attained the second-best score in $OMS_{ss}$ (39.79), highlighting the effectiveness of closed LLMs.

To validate the significance of these differences, we performed a two-sided pairwise Wilcoxon [79] signed-rank statistical test between $HelpCOM_N$ (GPT-4o) and each of the 12 baseline approaches across all evaluation metrics. In Table 4, an asterisk (*) is placed before a score if the difference between $HelpCOM_N$ (GPT-4o) and the corresponding approach is statistically significant (p-value < 0.05) for that metric. The results in Table 4 indicate that $HelpCOM_N$ (GPT-4o) demonstrates significant improvements across multiple metrics compared to the baselines. These findings support our hypothesis that incorporating helper method information enhances the quality of generated comments for dependent methods.

**Summary RQ₃:** $HelpCOM_N$ (GPT-4o) significantly outperformed the baselines, with improvements ranging from 5.6% to 50.4% across the evaluation metrics. This demonstrates that incorporating helper method information enhances the quality of generated comments for dependent methods.

## 4 Discussion
### 4.1 Interpreting the Generated Comments

We present an illustrative example in Figure 4 to highlight the effectiveness of HelpCOM. Specifically, we generated comments for the method *stopRandomDataNode()*, which relies on the helper method *ensureOpen()*. The summary shown is generated by $HelpCOM_N$ (GPT-4o). Compared to the baselines, HelpCOM produces a more logical and informative comment that better aligns with the ground truth. Notably, HelpCOM successfully incorporates insights from the *ensureOpen()* method, demonstrating its capability to generate information-rich summaries.

In contrast, CodeBERT and CodeT5+ generate comments that closely resemble the method name, overlooking its semantics. Similarly, ASAP produces a concise but less detailed summary compared to the ground-truth reference. Overall, baseline models struggle to generate high-quality comments, whereas HelpCOM effectively utilizes helper methods to produce more meaningful summaries.

### 4.2 Language Independent

The architecture of HelpCOM is designed to be language-independent. To evaluate its effectiveness across different programming languages, we selected one GitHub project each for Python[19] and PHP[20] and extracted all methods and their corresponding helper methods, following the same approach used for Java projects (Section 2.4). We then randomly sampled 50 dependent methods from each project and generated summaries using $HelpCOM_N$ (GPT-4o). The results are presented in Table 5.

---
[19]https://github.com/celery/celery
[20]https://github.com/sebastianbergmann/phpunit

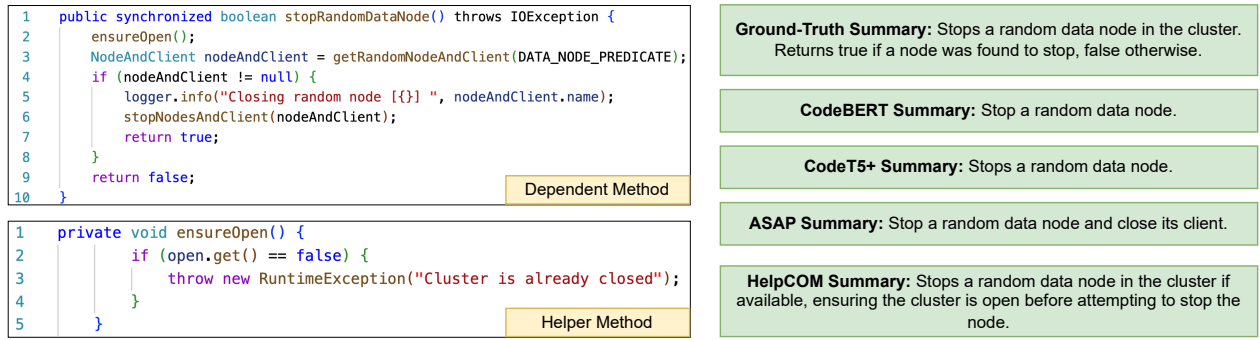

Figure 4: Qualitative analysis of the generated summaries Between HelpCOM and baselines.

Table 5: Evaluating HelpCOM: Python and PHP.

| Evaluation Metrics | | Python | PHP |
|---|---|---|---|
| Syntactic Similarity | BLEU | 38.18 | 16.03 |
| | METEOR | 70.63 | 37.81 |
| | ROUGE-L | 56.88 | 45.24 |
| Semantic Similarity | CIDEr | 46.19 | 18.61 |
| | SBERT | 79.22 | 71.91 |
| | USEnc | 74.98 | 67.59 |
| | SIDE | 88.08 | 85.16 |
| LLM-based | GPT-4 | 94.20 | 99.60 |
| | Llama-3.3 | 87.20 | 99.60 |
| $OMS_{ss}$ | | 64.35 | 48.03 |
| $OMS_{ssl}$ | | 73.56 | 66.05 |

From the results, we observe that HelpCOM performed well for Python ($OMS_{ssl}$ = 73.56) compared to PHP ($OMS_{ssl}$ = 66.05). Notably, HelpCOM achieved a higher overall metric score for both PHP and Python than for Java ($OMS_{ssl}$ = 56.82). While this could be attributed to the smaller sample size, the results also reinforce the language independence of HelpCOM.

### 4.3 Practitioner Survey Results

To evaluate the importance of helper methods information in the generation of code comments for dependent methods and check the comprehensibility of HelpCOM generated comments, we conducted an online developer survey hosted in SurveyMonkey[21] platform. In total, we received 162 responses, and among them, six responses were incomplete. Therefore, we continue our further analysis with the 156 complete responses.

We received participants from diverse backgrounds, with the majority (65.38%) being professional software developers, followed by students (14.10%), researchers (12.82%), and others (7.69%), including project managers, DevOps engineers, and software architects. In terms of programming experience, 40.38% had more than eight years of experience, 30.77% had 3–5 years, 15.38% had 6–8 years, and 13.46% had 0–2 years of experience, reflecting a well-balanced mix of expertise levels. Participants were also geographically diverse, representing 29 countries. The largest groups were from Bangladesh (28.85%), Canada (18.59%), Germany (8.33%), and the United States (12.82%), followed by Sweden, Australia, Italy, and the Netherlands.

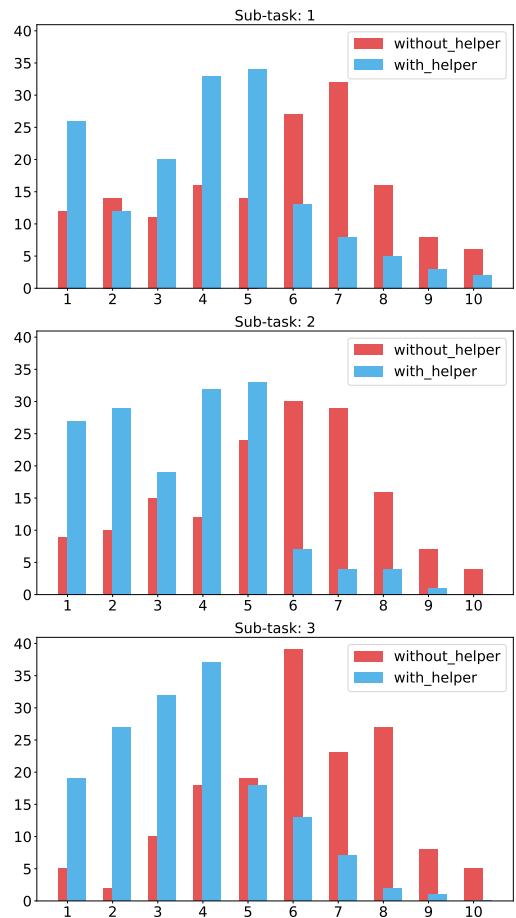

Figure 5: Practitioner understandability of given dependent method without and with helper method definition. X-axis and Y-axis present the rating (1-10) and frequency, respectively.

---
[21]https://www.surveymonkey.com/

Figure 5 illustrates the understandability levels of the three tasks presented in our online survey. As described in Section 2.8, each task was divided into three subtasks, where participants were asked to comprehend the dependent logic both with and without access to helper method definitions. The results reveal that participants consistently found the dependent method more challenging to understand when helper method definitions were hidden. The average scores for Task 1, Task 2, and Task 3 were (5.47, 4.06), (5.49, 3.50), and (6.02, 3.58), respectively, without and with helper method definitions. These results strongly support the motivation of our study, which suggests that automatic comment generation for dependent methods could significantly reduce the difficulty of understanding, especially when the generator takes into account the definitions of the helper methods it calls.

In Subtask 3, participants were asked to choose the most comprehensive comment based on the given code and helper method definitions. The comments were generated by the tools ASAP, CodeT5+, and our HelpCOM. Although participants were unaware of which tool generated which comment, approximately 75% of participants selected the comments generated by HelpCOM, followed by ASAP (18.38%) and CodeT5+ (7.05%). This result highlights the effectiveness of our proposed approach in comprehensive method-level comment generation. In addition, we asked participants whether receiving the comment earlier in subtask 1 would have helped them better understand the dependent method with less effort. Over 70% of participants agreed that the comment would have been helpful, while 16.67% remained neutral, and 13.25% felt that the comment would not have been useful.

We concluded the survey by asking participants for any additional feedback through an optional open-ended question. Some participants noted that well-crafted comments can help others understand code faster, especially when the comments reflect the structure or logic of the code. Example comments included: *"Yes, I would be interested in tools that generate comments based on context,"* and *"If the tool leveraged generative AI to infer comments based on the structure or logic of the code, that could be helpful."* However, concerns were raised about balancing helpful details with avoiding information overload. Participants emphasized that well-written code should be largely self-explanatory. Overall, the survey findings highlight the importance of generating context-aware and concise comments that improve code understandability without adding unnecessary complexity.

### 4.4 Implications

**For Researchers.** Our experiments ($RQ_3$) demonstrate that HelpCOM generates improved comments for dependent methods by incorporating information from helper methods, outperforming baseline techniques. Researchers working on code summarization should consider examining their approaches in greater depth to identify ways to enhance comment quality for dependent methods. Future research should build on our findings, striving to develop techniques that account for helper methods when generating comments.

**For Software Maintainers.** Our findings ($RQ_2$) reveal that existing automated comment-generation techniques often fail to produce high-quality comments for dependent methods, which make up a substantial portion of code in software repositories. This creates challenges for software maintainers when choosing tools for documenting uncommented projects, as semantically weak comments may confuse novice developers, increasing program comprehension time. To address this, software maintainers should prioritize validating the quality of generated comments when using comment-generation tools.

**For Novice Developers.** Since program comprehension is time-intensive [49, 84], novice developers may turn to automated comment-generation tools to streamline the process. While these tools can provide an overview of a method's purpose, they may also inadvertently lead developers in the wrong direction ($RQ_2$). In cases of ambiguity, novice developers should seek guidance from experienced colleagues to ensure accurate understanding.

## 5 Related Work
### 5.1 Automatic Code Comment Generation

The evolution of automatic code comment generation has seen significant advancements over the years. Early methods primarily relied on template-based approaches [25, 65] and information retrieval (IR) [15, 24, 59], which were limited in flexibility and accuracy. Template-based approaches struggled with achieving generalization, as they required predefined structures to generate comments for code snippets. Similarly, IR-based methods retrieved existing code-comment pairs based on similarity, but this approach was dependent on the availability of sufficiently similar examples in the dataset.

As deep learning gained traction, researchers began applying neural networks, such as neural machine translation (NMT) models, to automatically generate comments. Iyer et al. [32] were among the first to treat comment generation as an end-to-end translation task, adapting NMT techniques for this purpose. This shift marked a significant departure from traditional methods, as neural models could learn directly from large-scale code-comment corpora, producing more contextually relevant and accurate summaries. The introduction of the seq2seq [68] model for code comment generation by Zhang et al. [86] further improved this approach by leveraging similar code snippets to generate higher-quality comments. Techniques like AST-Trans [69], which utilized the abstract syntax tree (AST) structure, further refined the comment generation process by incorporating hierarchical relationships in the code.

With the rise of pre-trained transformer [71] models, such as CodeBERT [16] and CodeT5 [77], the focus shifted to utilizing these large models fine-tuned on code summarization tasks. These transformer-based models, trained on vast code datasets, demonstrated significant improvements in performance. However, these models still required substantial fine-tuning and the creation of language-specific versions, leading to high training costs and time. The introduction of multilingual fine-tuning helped mitigate this problem, but it did not eliminate the challenges associated with training time and data dependency.

The latest wave of advancements has been driven by LLMs, such as GPT-based models, which are pre-trained on vast datasets and exhibit impressive zero-shot [33] and few-shot [58] learning capabilities. LLMs can generate accurate comments with minimal fine-tuning, simply by being provided with a few exemplars or prompts, thus eliminating the need for extensive labeled data. This has made LLMs more efficient and flexible, leading to significant improvements in automatic comment-generation tasks. Studies [4, 20] have shown that LLMs outperform smaller pre-trained models and fine-tuned transformers, particularly when applied to code summarization and comment generation tasks, by providing high-quality, contextually relevant results even in complex scenarios.

### 5.2 Reasoning with LLMs

Human reasoning encompasses critical thinking, detailed information analysis, and decision-making grounded in prior experience and contextual insight. Studies have assessed LLMs' reasoning abilities, showing considerable progress in this area [14, 56]. Consequently, LLMs are increasingly applied across software engineering domains, including code generation [7, 11, 22, 53, 85], program repair [82, 83], bug detection [38, 43], code review [41, 45], and requirements analysis [34, 48]. Motivated by these reasoning capabilities, we incorporate an LLM into HelpCOM's architecture.

With advancements in LLMs, in-context learning (ICL) has become a key NLP paradigm, where LLMs make predictions based on prompts supplemented by relevant examples [8, 14]. In HelpCOM, we utilize ICL by structuring prompts to include helper methods, enhancing the quality of generated comments.

## 6 Threats to Validity

While we have made every effort to ensure accuracy, potential validity threats may still impact the study's results. Following Runeson et al.'s classification [62], we examine several factors that could affect the study's validity.

**Construct Validity.** Construct validity assesses how well the study measures the intended concepts. In this study, we based our approach on the assumption that dependent and independent methods differ significantly and that incorporating helper methods would enhance comment quality for dependent methods. To validate this, we conducted a quantitative analysis showing that dependent methods play a substantial role in software repositories ($RQ_1$) and achieve better results when helper methods are utilized ($RQ_3$). The prompt designed to generate method comments may also introduce validity risks. To mitigate this, we followed OpenAI's prompt structuring guidelines and referenced related studies [4, 20, 52].

**Internal Validity.** HelpCOM utilizes GPT-4o, which may have encountered similar methods and ground-truth comments from GitHub during training, as these models often draw on open-source data. This could introduce bias into our results; however, our experiments show that GPT-4o, without including helper method information, performed poorly compared to HelpCOM. The context window of GPT-4o is 128K tokens, and none of the prompts in our experiment exceeded it. On average, we encountered two levels of helpers per dependent method, which indicates that a 128K token size is sufficient. Due to API cost constraints, we evaluated HelpCOM and other baselines on a subset of the dataset, following the approach used in related studies [4, 67]. To minimize selection bias, we used random sampling with a 95% confidence level and a 5% margin of error, following standard statistical methods [51].

**External Validity.** Our study used only Java projects, which may impact the generalizability of the findings to other programming languages. However, after experimenting with one Python project and one PHP project, we found that HelpCOM is independent of programming language (Table 5). A limitation of using an LLM in software research is its non-deterministic nature [54], which can produce different outputs with repeated attempts. To mitigate this and increase consistency in HelpCOM's output, we set the temperature to 0.2, which enhances determinism when generating comments from GPT-4o.

**Conclusion Validity.** We present the results of our experiments using various evaluation metrics, including syntactic (e.g., BLEU), semantic (e.g., SentenceBERT), LLM-based metrics, and overall score metrics. Our findings demonstrate that baseline models struggle to generate comments for dependent methods, unlike for independent methods ($RQ_2$). Additionally, we show that HelpCOM outperforms the baselines across these metrics ($RQ_3$). To confirm the statistical significance of the observed differences, we performed a two-sided pairwise Wilcoxon signed-rank test, which yielded significant p-values ($< 0.05$).

## 7 Conclusion and Future Work

In this study, we proposed a noble technique called HelpCOM, which leverages the helper methods of a dependent method to generate high-quality comments using the capabilities of different LLMs (e.g., GPT-4o). We created a dataset consisting of dependent and helper methods from 10 Java projects on GitHub. First, we demonstrated that dependent methods are not insignificant, as they represent a substantial portion (69.25%) of methods in software repositories. Next, we evaluated the baseline comment generation techniques thoroughly and found out they perform poorly when generating comments for dependent methods, compared to independent methods. Finally, we compared HelpCOM's performance against the baselines and found that HelpCOM outperforms them in evaluation metrics, with a 5.6% to 50.4% increase in Overall Metric Score ($OMS_{ssl}$), demonstrating its ability to generate higher-quality comments for dependent methods. Additionally, a comprehensive user survey with 156 software developers evaluated the quality of HelpCOM, with participants agreeing on its potential to automatically generate method-level comments that are highly comprehensible. For future work, we aim to develop a tool that utilizes HelpCOM to generate better comments, thereby aiding program comprehension, as suggested by the participants in our survey.

## Acknowledgments

This research is supported in part by the Natural Sciences and Engineering Research Council of Canada (NSERC) Discovery Grants program, the Canada Foundation for Innovation's John R. Evans Leaders Fund (CFI-JELF), and by the industry-stream NSERC CREATE in Software Analytics Research (SOAR).